\documentclass{Interspeech}
\usepackage{cite,colortbl,tabularx}
\hypersetup{colorlinks=true,citecolor=green,linkcolor=red,urlcolor=magenta}

\newcommand{\better}[1]{\multicolumn{1}{>{\columncolor[rgb]{1.0,0.85,0.85}}c}{#1}}
\newcommand{\goodp}[1]{\multicolumn{1}{>{\columncolor[rgb]{1.0,0.87,0.87}}c}{#1}}
\newcommand{\good}[1]{\multicolumn{1}{>{\columncolor[rgb]{1.0,0.9,0.9}}c}{#1}}
\newcommand{\goodm}[1]{\multicolumn{1}{>{\columncolor[rgb]{1.0,0.95,0.95}}c}{#1}}
\newcommand{\worstpp}[1]{\multicolumn{1}{>{\columncolor[rgb]{0.7,0.7,1.0}}c}{#1}}
\newcommand{\worstp}[1]{\multicolumn{1}{>{\columncolor[rgb]{0.75,0.75,1.0}}c}{#1}}
\newcommand{\worst}[1]{\multicolumn{1}{>{\columncolor[rgb]{0.8,0.8,1.0}}c}{#1}}
\newcommand{\worse}[1]{\multicolumn{1}{>{\columncolor[rgb]{0.85,0.85,1.0}}c}{#1}}
\newcommand{\bad}[1]{\multicolumn{1}{>{\columncolor[rgb]{0.9,0.9,1.0}}c}{#1}}
\newcommand{\badm}[1]{\multicolumn{1}{>{\columncolor[rgb]{0.95,0.95,1.0}}c}{#1}}
\newcommand{\customparagraph}[1]{\noindent\textbf{#1}}
\newcommand{\customsubparagraph}[1]{\noindent\textit{#1}}
\newcolumntype{C}{>{\centering\arraybackslash}X}
\makeatletter
\newcommand\footnoteref[1]{\protected@xdef\@thefnmark{\ref{#1}}\@footnotemark}
\makeatother

\interspeechcameraready

\title{Vocoder-Projected Feature Discriminator}

\author{Takuhiro}{Kaneko}
\author{Hirokazu}{Kameoka}
\author{Kou}{Tanaka}
\author{Yuto}{Kondo}

\affiliation[nocounter]{}{NTT, Inc.}{Japan}
\email{takuhiro.kaneko@ntt.com}

\begin{document}

\maketitle

\begin{abstract}
  In text-to-speech (TTS) and voice conversion (VC), acoustic features, such as mel spectrograms, are typically used as synthesis or conversion targets owing to their compactness and ease of learning. However, because the ultimate goal is to generate high-quality waveforms, employing a vocoder to convert these features into waveforms and applying adversarial training in the time domain is reasonable. Nevertheless, upsampling the waveform introduces significant time and memory overheads. To address this issue, we propose a \textit{vocoder-projected feature discriminator (VPFD)}, which uses vocoder features for adversarial training. Experiments on diffusion-based VC distillation demonstrated that a pretrained and frozen vocoder feature extractor with a single upsampling step is necessary and sufficient to achieve a VC performance comparable to that of waveform discriminators while reducing the training time and memory consumption by 9.6 and 11.4 times, respectively.\footnote{\label{foot:samples}Audio samples are available at \url{https://www.kecl.ntt.co.jp/people/kaneko.takuhiro/projects/vpfd/}.}
\end{abstract}

\section{Introduction}
\label{sec:introduction}

Text-to-speech (TTS) and voice conversion (VC) are techniques designed to generate speech from text and speech inputs, respectively.
In TTS and VC, a widely adopted approach is the two-stage framework, where the first model generates acoustic features (e.g., mel spectrograms) from input data (e.g., text or acoustic features) and the second model, called the vocoder, synthesizes the waveform from the generated acoustic features.
Compared to an end-to-end approach, the two-stage approach offers advantages such as more compact learning and greater portability of individual modules, which have led to intensive research.

\begin{figure}[t]  
  \centering
  \includegraphics[width=0.99\linewidth]{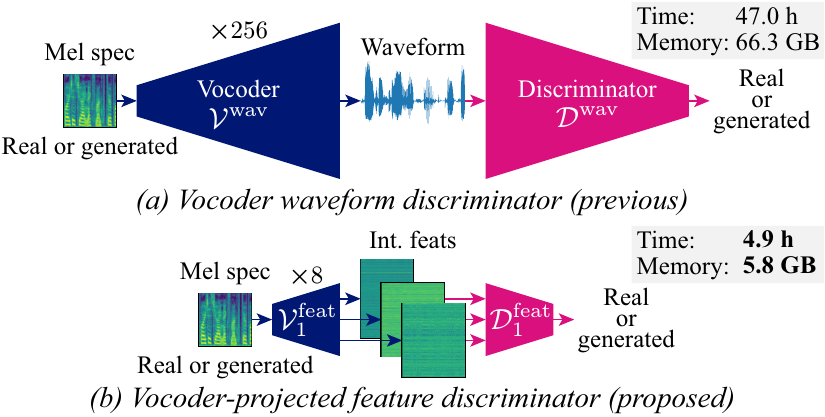}
  \vspace{-2mm}
  \caption{Comparison between (a) vocoder waveform discriminator (VWD, previous) and (b) vocoder-projected feature discriminator (\textit{VPFD}, proposed).
    In \textit{VPFD}, the discriminator is constructed using intermediate features (int. feats) from a pretrained vocoder, thereby bypassing the need to learn an effective representation for waveform synthesis and reducing training time and memory consumption.}
  \label{fig:teaser}
  \vspace{-4mm}
\end{figure}

Realistic acoustic features must be generated to synthesize high-quality speech.
Generative adversarial network (GAN)~\cite{IGoodfellowNIPS2014}-based adversarial training, where an acoustic feature generator is trained adversarially with a discriminator, has been widely adopted to achieve this objective (e.g.,~\cite{TKanekoICASSP2017,TKanekoIS2017a,YSaitoTASLP2017,CHsuIS2017,TKanekoEUSIPCO2018,HKameokaSLT2018,TKanekoIS2019,HKameokaTASLP2020c,TKanekoICASSP2021,YLiIS2021,TKanekoIS2024}).
Specifically, considering that the ultimate goal is to synthesize high-quality waveforms, it is reasonable to focus on improving the quality of the generated acoustic features in the time domain.
To achieve this objective, a previous study~\cite{TKanekoIS2024} proposed the vocoder waveform discriminator (VWD), which converts acoustic features into a waveform using a vocoder (e.g.,~\cite{KKumarNeurIPS2019,RYamamotoICASSP2020,JKongNeurIPS2020,GYangSLT2021,WJangIS2021,TKanekoICASSP2022,SLeeICLR2023,TKanekoIS2023}) and subsequently distinguishes between the real and synthesized waveforms using a waveform discriminator (e.g.,~\cite{KKumarNeurIPS2019,JKongNeurIPS2020,WJangIS2021,TKanekoICASSP2023}), as shown in Figure~\ref{fig:teaser}(a).
The effectiveness of VWD has been demonstrated, for example, in diffusion-based VC distillation~\cite{TKanekoIS2024}, where it enhances speech quality and speaker similarity while stabilizing GAN training---achievements that were challenging using typical mel-spectrogram discriminators.
This can be attributed to the fact that mel-spectrogram discriminators must not only learn to distinguish between real and generated data but also capture a representation that reflects the reality of the waveform.
VWD alleviates the latter requirement by employing a pretrained vocoder, thereby facilitating the aforementioned benefits.
However, the limitation of this approach is that upsampling from acoustic features to the waveform, e.g., by 256 times, introduces time and memory overheads.
Consequently, training is infeasible under conditions with limited time and computational resources.

To overcome this limitation, we propose a novel discriminator called the \textit{vocoder-projected feature discriminator (VPFD)}, which distinguishes between real and generated data using the intermediate features of the vocoder rather than the waveform obtained from the output of the vocoder, as shown in Figure~\ref{fig:teaser}(b).
The core idea is to reduce the computation time and memory consumption by utilizing intermediate features derived from fewer upsampling steps (e.g., 8 times) instead of the waveform obtained through full upsampling (e.g., 256 times).

The two key questions in this approach are as follows: \textit{Q1. To what extent can upsampling be reduced?} \textit{Q2. How should the vocoder feature extractor be handled during training?}
In particular, regarding the latter, inspired by Projected GAN~\cite{ASauerNeurIPS2021,ASauerSIGGRAPH2022,ASauerICML2023}, we explored the importance of pretraining and freezing parameters.
In our experiments, we addressed these questions by applying \textit{VPFD} to diffusion-based VC distillation~\cite{TKanekoIS2024}.
The results show that a pretrained and frozen vocoder with a single upsampling is necessary and sufficient to achieve VC performance comparable to that obtained with VWD.
By reducing upsampling, \textit{VPFD} reduces the training time and memory consumption by factors of 9.6 and 11.4, respectively.
Improving the quality of acoustic features is essential for various tasks.
Therefore, the findings of this study are expected to have broader applications.

The remainder of this paper is organized as follows.
Section~\ref{sec:method} outlines the problem setting, describes the baseline VWD, and introduces the proposed \textit{VPFD} and its application to diffusion-based VC distillation.
Section~\ref{sec:experiments} presents the experimental results.
Finally, Section~\ref{sec:conclusion} concludes the paper and discusses potential future research.

\section{Method}
\label{sec:method}

\subsection{Problem setting}
\label{subsec:problem}

As discussed in Section~\ref{sec:introduction}, this study focused on the acoustic feature generator $\mathcal{G}$ (first-stage model), which generates acoustic features $\bm{x}^g$ from the input data $\bm{z}$ (e.g., text or acoustic features) within the two-stage framework of TTS and VC.
The goal is to enhance the quality of $\bm{x}^g = \mathcal{G}(\bm{z})$, ensuring that it is comparable to the real acoustic features $\bm{x}^r$ in the training set.
To achieve this, we employed GAN~\cite{IGoodfellowNIPS2014}-based adversarial training.
In Sections~\ref{subsec:preliminary} and \ref{subsec:proposal}, we discuss the components of the previous VWD~\cite{TKanekoIS2024} and proposed \textit{VPFD}, both of which are designed to achieve this objective.

\subsection{Preliminary: Vocoder waveform discriminator}
\label{subsec:preliminary}

In the above-mentioned problem, the ultimate goal is to generate high-quality waveforms; therefore, it is reasonable to improve the realism of acoustic features in the time domain.
To achieve this objective, VWD~\cite{TKanekoIS2024} converts $\bm{x}^r$ and $\bm{x}^g$ into waveforms using a vocoder $\mathcal{V}^{\textnormal{wav}}$ (e.g.,~\cite{KKumarNeurIPS2019,RYamamotoICASSP2020,JKongNeurIPS2020,GYangSLT2021,WJangIS2021,TKanekoICASSP2022,SLeeICLR2023,TKanekoIS2023}) and distinguishes them in the time domain using a waveform discriminator $\mathcal{D}^{\textnormal{wav}}$ (e.g.,~\cite{KKumarNeurIPS2019,JKongNeurIPS2020,WJangIS2021,TKanekoICASSP2023}).
The loss function (specifically, the least squares GAN form~\cite{XMaoICCV2017}) is expressed as
\begin{flalign}
  \label{eq:adv_loss_d_vwd}
  \mathcal{L}_{\mathcal{D}}^{\textnormal{VWD}}
  & = \mathbb{E}_{\bm{x}^r} (\mathcal{D}^{\textnormal{wav}}(\mathcal{V}^{\textnormal{wav}}(\bm{x}^r)) - 1)^2
  + \mathbb{E}_{\bm{z}} (\mathcal{D}^{\textnormal{wav}}(\mathcal{V}^{\textnormal{wav}}(\bm{x}^g)))^2,
  \\
  \label{eq:adv_loss_g_vwd}
  \mathcal{L}_{\mathcal{G}}^{\textnormal{VWD}}
  & = \mathbb{E}_{\bm{z}} (\mathcal{D}^{\textnormal{wav}}(\mathcal{V}^{\textnormal{wav}}(\bm{x}^g)) - 1)^2,
\end{flalign}
where $\mathcal{D}^{\textnormal{wav}}$ attempts to distinguish $\bm{x}^r$ and $\bm{x}^g$ in the time domain by minimizing $\mathcal{L}_{\mathcal{D}}^{\textnormal{VWD}}$, whereas $\mathcal{G}$ attempts to generate $\bm{x}^g$ that can deceive $\mathcal{D}^{\textnormal{wav}}$ in the time domain by minimizing $\mathcal{L}_{\mathcal{G}}^{\textnormal{VWD}}$.
To stabilize the GAN training, the feature matching (FM) loss~\cite{ALarsenICML2016,KKumarNeurIPS2019,KOyamadaEUSIPCO2018} is also commonly used as follows:
\begin{flalign}
  \label{eq:fm_loss_vwd}
  \mathcal{L}_{\textnormal{FM}}^{\textnormal{VWD}}
  = \mathbb{E}_{\bm{x}^r, \bm{z}} \sum_{i=1}^{M^{\textnormal{wav}}} \frac{1}{N_i^{\textnormal{wav}}} \lVert \mathcal{D}_i^{\textnormal{wav}}(\mathcal{V}^{\textnormal{wav}}(\bm{x}^r)) - \mathcal{D}_i^{\textnormal{wav}}(\mathcal{V}^{\textnormal{wav}}(\bm{x}^g)) \rVert_1,
\end{flalign}
where $M^{\textnormal{wav}}$ is the number of layers in $\mathcal{D}^{\textnormal{wav}}$, and $\mathcal{D}_i^{\textnormal{wav}}$ and $N_i^{\textnormal{wav}}$ denote the features and number of features in the $i$-th layer of $\mathcal{D}^{\textnormal{wav}}$, respectively.
This loss encourages $\bm{x}^g$ to be closer to $\bm{x}^r$ in the feature space of $\mathcal{D}^{\textnormal{wav}}$.

\subsection{Proposal: Vocoder-projected feature discriminator}
\label{subsec:proposal}

As discussed in Section~\ref{sec:introduction}, VWD improves the quality of the synthesized waveform and stabilizes GAN training.
However, upsampling from acoustic features to the waveform (e.g., 256 times) significantly increases training time and memory usage.
To overcome this limitation while retaining the advantages of VWD, that is, bypassing the need to learn an effective representation for waveform synthesis, we propose \textit{VPFD}, which distinguishes between $\bm{x}^r$ and $\bm{x}^g$ using the intermediate features of a pretrained vocoder $\mathcal{V}^{\textnormal{wav}}$.
Let the intermediate feature extractor up to the $L$-th upsampling layer of the vocoder be denoted by $\mathcal{V}_L^{\textnormal{feat}}$, and let the discriminator that performs discrimination using these features be denoted by $\mathcal{D}_L^{\textnormal{feat}}$.
The more the upsampling steps are reduced, the more the training time and memory consumption can be reduced.
The necessary number of upsampling steps, i.e., the setting of $L$, is experimentally evaluated in Section~\ref{subsec:ablation_study}.
The training objective and network architecture of \textit{VPFD} are detailed below.

\smallskip
\customparagraph{Training objective.}
In \textit{VPFD}, the adversarial loss (Equations~\ref{eq:adv_loss_d_vwd} and \ref{eq:adv_loss_g_vwd}) and FM loss (Equation~\ref{eq:fm_loss_vwd}) are redefined as follows:
\begin{flalign}
  \label{eq:adv_loss_d_vpfd}
  \mathcal{L}_{\mathcal{D}}^{\textnormal{VPFD}_L}
  & = \mathbb{E}_{\bm{x}^r} (\mathcal{D}_L^{\textnormal{feat}}(\mathcal{V}_L^{\textnormal{feat}}(\bm{x}^r)) - 1)^2
  + \mathbb{E}_{\bm{z}} (\mathcal{D}_L^{\textnormal{feat}}(\mathcal{V}_L^{\textnormal{feat}}(\bm{x}^g)))^2,
  \\
  \label{eq:adv_loss_g_vpfd}
  \mathcal{L}_{\mathcal{G}}^{\textnormal{VPFD}_L}
  & = \mathbb{E}_{\bm{z}} (\mathcal{D}_L^{\textnormal{feat}}(\mathcal{V}_L^{\textnormal{feat}}(\bm{x}^g)) - 1)^2,
  \\
  \label{eq:fm_loss_vpfd}
  \mathcal{L}_{\textnormal{FM}}^{\textnormal{VPFD}_L}
  & = \mathbb{E}_{\bm{x}^r, \bm{z}} \sum_{i=1}^{M_L^{\textnormal{feat}}} \frac{1}{N_{L, i}^{\textnormal{feat}}} \lVert \mathcal{D}_{L, i}^{\textnormal{feat}}(\mathcal{V}_L^{\textnormal{feat}}(\bm{x}^r)) - \mathcal{D}_{L, i}^{\textnormal{feat}}(\mathcal{V}_L^{\textnormal{feat}}(\bm{x}^g)) \rVert_1,
\end{flalign}
where $M_L^{\textnormal{feat}}$ is the number of layers in $\mathcal{D}_L^{\textnormal{feat}}$, and $\mathcal{D}_{L, i}^{\textnormal{feat}}$ and $N_{L, i}^{\textnormal{feat}}$ denote the features and number of features in the $i$-th layer of $\mathcal{D}_L^{\textnormal{feat}}$, respectively.

\begin{figure}[t]  
  \centering
  \includegraphics[width=0.95\linewidth]{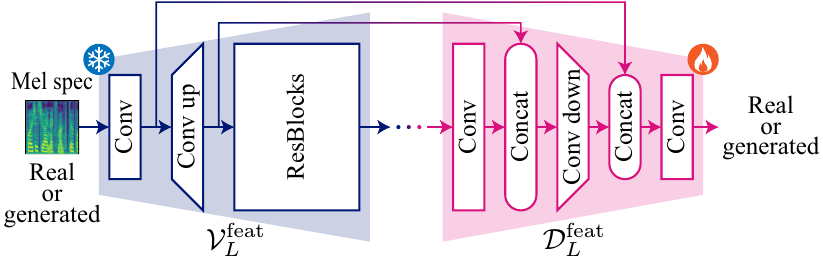}
  \vspace{-2mm}
  \caption{Network architecture of vocoder-projected feature discriminator (\textit{VPFD}).
    \textit{VPFD} is composed of $\mathcal{V}_L^{\textnormal{feat}}$ and $\mathcal{D}_L^{\textnormal{feat}}$.
    \texttt{Conv}, \texttt{ResBlocks}, and \texttt{Concat} indicate a convolution layer, residual blocks~\cite{KHeCVPR2016}, and a channel concatenation operation, respectively.}
  \label{fig:network}
  \vspace{-4mm}
\end{figure}

\smallskip
\customparagraph{Network architecture.}
Based on the finding for Projected GAN~\cite{ASauerNeurIPS2021,ASauerSIGGRAPH2022,ASauerICML2023} that multiscale feature extraction is effective, we employ a network architecture based on U-Net~\cite{ORonnebergerMICCAI2015}, as shown in Figure~\ref{fig:network}.\footnote{More specifically, we employ an ``inverted'' U-Net.
  While the original U-Net~\cite{ORonnebergerMICCAI2015} utilizes a structure that performs downsampling followed by upsampling, it should be noted that in this case, we use a structure that performs upsampling followed by downsampling.}
In $\mathcal{D}_L^{feat}$, downsampling is performed at the same rate as upsampling in $\mathcal{V}_L^{feat}$.
During downsampling, the kernel size is set to twice the downsampling rate, while in other cases, it is set to $21$.
The number of output channels is set to equal the number of layers in $\mathcal{V}_L^{feat}$ at the same scale.
A leaky rectified linear unit~\cite{AMaasICML2013} is used as the activation function, and weight normalization~\cite{TSalimansNIPS2016} is applied to all the convolution layers.
In the default setting, $\mathcal{V}_L^{\textnormal{feat}}$ is pretrained and its parameters are frozen during training.
In contrast, the parameters of $\mathcal{D}_L^{\textnormal{feat}}$ are optimized during training.
The effectiveness of this strategy is experimentally validated in Section~\ref{subsec:ablation_study}.

\subsection{Application to distillation of diffusion-based VC}
\label{subsec:application}

In the experiments (Section~\ref{sec:experiments}), we validated \textit{VPFD} by applying it to the adversarial distillation of diffusion-based VC~\cite{TKanekoIS2024}.
In this distillation, VoiceGrad~\cite{HKameokaTASLP2024}, a denoising diffusion probabilistic model (DDPM)~\cite{JHoNeurIPS2020}-based nonparallel VC, is distilled into a one-step diffusion-based VC (FastVoiceGrad; FVG) by leveraging both GANs~\cite{IGoodfellowNIPS2014} and diffusion models~\cite{JSohlICML2015}.
We evaluated the performance when VWD, which was originally used in FVG, was replaced with \textit{VPFD}.
In this section, we first provide a brief overview of VoiceGrad, followed by a description of FVG configured using \textit{VPFD}.

\smallskip
\customparagraph{VoiceGrad.}
VoiceGrad is a nonparallel VC model that generates $\bm{x}^g$ from $\bm{x}^r$, conditioned on a speaker embedding $\bm{s}$~\cite{YJiaNeurIPS2018} and content embedding $\bm{p}$~\cite{SLiuTASPL2021}, using diffusion and reverse diffusion processes.
During training, reconstruction is performed using $\bm{x}^r$, $\bm{s}$, and $\bm{p}$ extracted from the speech of the same speaker, whereas during inference, conversion is performed using $\bm{x}^r$ and $\bm{p}$ extracted from the source speaker's speech and $\bm{s}$ extracted from the target speaker's speech.
The diffusion and reverse diffusion processes are described below.

\smallskip
\customsubparagraph{Diffusion process.}
During this process, $\bm{x}^r$ ($= \bm{x}_0$) is gradually transformed into noise $\bm{x}_T \sim \mathcal{N}(\bm{0}, \bm{I})$ over $T$ steps (where $T = 1000$ in practice).
Owing to the reproductivity of the normal distribution and a reparameterization trick~\cite{DKingmaICLR2014}, the $t$-step diffused data $\bm{x}_t$ ($t \in \{1, \dots, T \}$) are expressed as
\begin{flalign}
  \label{eq:diffusion_process}
  \bm{x}_t = \sqrt{\bar{\alpha}_t} \bm{x}_0 + \sqrt{1 - \bar{\alpha}_t} \bm{\epsilon},
\end{flalign}
where $\bar{\alpha} = \prod_{i = 1}^t \alpha_t$, $1 - \alpha_i$ represents the noise variance at the $i$-th step, and $\bm{\epsilon} \sim \mathcal{N}(\bm{0}, \bm{I})$.

\smallskip
\customsubparagraph{Reverse diffusion process.}
During this process, $\bm{x}_t$ is gradually denoised towards $\bm{x}_0$.
This denoising process is given by
\begin{flalign}
  \label{eq:reverse_diffusion_process}
  \hspace{-2mm}
  \bm{\mu}_{\theta} (\bm{x}, t, \bm{s}, \bm{p})
  = \frac{1}{\sqrt{\alpha_t}} \left( \bm{x}_t - \frac{1 - \alpha_t}{\sqrt{1 - \bar{\alpha}_t}} \bm{\epsilon}_{\theta} (\bm{x}_t, t, \bm{s}, \bm{p}) \right),
\end{flalign}
where $\bm{\epsilon}_{\theta}$ is a denoising function parameterized by $\theta$.

\smallskip
\customparagraph{FVG with VPFD.}
In FVG, a pretrained VoiceGrad model is used as an initial condition, and its denoising process (i.e., Equation~\ref{eq:reverse_diffusion_process}) is distilled such that its output remains feasible even when this process is performed only once.
This is achieved by distilling the model using adversarial loss~\cite{IGoodfellowNIPS2014,XMaoICCV2017} and score distillation loss~\cite{ASauerECCV2024}.
For clarity, we denote the parameters of the student model (i.e., FVG) and teacher model (i.e., VoiceGrad) by $\phi$ and $\theta$, respectively.

\smallskip
\customsubparagraph{Adversarial loss.}
The adversarial loss (including the FM loss) configured with \textit{VPFD} is defined by Equations~\ref{eq:adv_loss_d_vpfd}--\ref{eq:fm_loss_vpfd}, where $\bm{x}^g$ is generated by $\bm{\mu}_{\phi}$ (calculated as in Equation~\ref{eq:reverse_diffusion_process}) and $\bm{z} = \bm{x}^r$.
We denote this $\bm{x}^g$ as $\bm{x}_{\phi}^g$.

\smallskip
\customsubparagraph{Score distillation loss.}
The score distillation loss is expressed as
\begin{flalign}
  \label{eq:dist_loss}
  \mathcal{L}_{\textnormal{distill}} = \mathbb{E}_{t, \bm{x}^r} \sqrt{\bar{\alpha}_t} \lVert \bm{x}_{\phi}^g - \bm{x}_{\theta}^g \rVert,
\end{flalign}
where $\bm{x}_{\theta}^g = \bm{\mu}_{\theta} (\textnormal{sg}(\bm{x}_{\phi, t}^g), t, \bm{s}, \bm{p})$, $\textnormal{sg}$ represents the stop gradient operation, $\bm{x}_{\phi, t}^g$ is the $t$-step diffused $\bm{x}_{\phi}^g$, and $t \in \{1, \dots, T \}$.
This loss encourages $\bm{x}_{\phi}^g$ to match the output obtained by diffusing and then reversely diffusing it using the teacher model $\bm{\mu}_{\theta}$.

\smallskip
\customsubparagraph{Total objective.}
The total objective function is expressed as
\begin{flalign}
  \label{eq:total_loss}
  \mathcal{L}_{\mathcal{D}} & = \mathcal{L}_{\mathcal{D}}^{\textnormal{VPFD}_L},
  \\
  \mathcal{L}_{\mathcal{G}} & = \mathcal{L}_{\mathcal{G}}^{\textnormal{VPFD}_L}
                              + \lambda_{\textnormal{FM}} \mathcal{L}_{\textnormal{FM}}^{\textnormal{VPFD}_L}
                              + \lambda_{\textnormal{distill}} \mathcal{L}_{\textnormal{distill}},
\end{flalign}
where $\lambda_{\textnormal{FM}}$ and $\lambda_{\textnormal{distill}}$ are weighting hyperparameters, set to $2$ and $45$, respectively, in the experiments, following the FVG study~\cite{TKanekoIS2024}.
$\mathcal{D}_L^{\textnormal{feat}}$ is optimized by minimizing $\mathcal{L}_{\mathcal{D}}$, whereas $\mathcal{G}$ is optimized by minimizing $\mathcal{L}_{\mathcal{G}}$.

\section{Experiments}
\label{sec:experiments}

\subsection{Experimental setup}
\label{subsec:experimental_setup}

\customparagraph{Data.}
The experimental setup related to the data followed that of the FVG study~\cite{TKanekoIS2024}.
We evaluated \textit{VPFD} for one-shot any-to-any VC.
The main experiments (Sections~\ref{subsec:ablation_study} and \ref{subsec:comparative_study}) were conducted using the VCTK dataset~\cite{JYamagishiVCTK2019}, which includes utterances from 110 English speakers.
Generalizability, independent of the dataset, was evaluated using the LibriTTS dataset~\cite{HZenIS2019}, which contains utterances from approximately 1,100 English speakers (Section~\ref{subsec:generalizability_analysis}).
For the unseen-to-unseen VC scenario, 10 speakers and 10 sentences were excluded for evaluation, while the remaining data were used for training.
The audio clips were downsampled to 22.05 kHz, from which 80-dimensional log-mel spectrograms were extracted with an FFT size of 1024, hop size of 256, and window size of 1024.
These log-mel spectrograms were used as conversion targets.

\smallskip
\customparagraph{Implementation.}
For a fair comparison, we implemented our model (\textit{FVG$+$VPFD$_L$}) based on FVG~\cite{TKanekoIS2024}, with the only modification being the replacement of the discriminator from VWD with \textit{VPFD$_L$}.
Both $\bm{\mu}_{\theta}$ and $\bm{\mu}_{\phi}$ were implemented using the U-Net~\cite{ORonnebergerMICCAI2015} architecture, consisting of 12 convolution layers with 512 hidden channels, two downsampling/upsampling processes, gated linear units~\cite{YDauphinICML2017}, and weight normalization~\cite{TSalimansNIPS2016}.
$\bm{s}$ was extracted using a speaker encoder~\cite{YJiaNeurIPS2018}, while $\bm{p}$ was obtained using a bottleneck feature extractor~\cite{SLiuTASPL2021}.
For $\mathcal{V}$, we utilized a pretrained HiFi-GAN V1 generator~\cite{JKongNeurIPS2020}.\footnote{\url{https://github.com/jik876/hifi-gan}}
The baseline VWD used $\mathcal{D}^{\textnormal{wav}}$, which combines the multiperiod discriminator (MPD)~\cite{JKongNeurIPS2020} and multiresolution discriminator (MRD)~\cite{WJangIS2021}.
In contrast, the proposed \textit{VPFD} used $\mathcal{D}^{\textnormal{feat}}$, implemented as shown in Figure~\ref{fig:network}.
The models were trained using the Adam optimizer~\cite{DPKingmaICLR2015} with a batch size of 32, learning rate of 0.0002, $\beta_1$ of 0.5, and $\beta_2$ of 0.9 for 100 epochs on VCTK and 50 epochs on LibriTTS.
The waveforms were synthesized using $\mathcal{V}$.

\smallskip
\customparagraph{Evaluation metrics.}
To efficiently investigate various models, we primarily employed objective metrics and conducted a subjective evaluation for the most critical comparison (Table~\ref{tab:subjective_evaluation}).
The following six objective metrics were used:
(1) \textit{UTMOS$\uparrow$}~\cite{TSaekiIS2022}: the predicted mean opinion score (MOS) is tuned to assess the quality of synthesized speech.
(2) \textit{DNSMOS$\uparrow$}~\cite{CReddyICASSP2021}: the predicted MOS is optimized to evaluate the quality of noise-suppressed speech.
(3) Character error rate (\textit{CER$\downarrow$}) by Whisper-large-v3\cite{ARadfordICML2023}: measures speech intelligibility.
(4) Speaker encoder cosine similarity (\textit{SECS$\uparrow$}) by Resemblyzer\footnote{\url{https://github.com/resemble-ai/Resemblyzer}}: quantifies speaker similarity.
(5) \textit{Time$\downarrow$}: training time (in hours) measured on a single A100 GPU.
(6) \textit{Memory$\downarrow$}: maximum memory consumption (in gigabytes) during training.
For each metric, $\uparrow$ indicates that a larger value is better, while $\downarrow$ indicates that a smaller value is better.

\subsection{Ablation study}
\label{subsec:ablation_study}

\begin{figure}[h]
  \centering
  \includegraphics[width=0.99\linewidth]{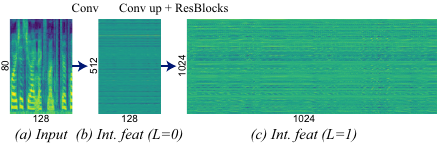}
  \vspace{-2mm}
  \caption{Comparison of input and intermediate features.
    \textbf{Best viewed when zoomed in}.
    The aspect is adjusted for clarity.}
  \label{fig:intfeat}
  \vspace{-3mm}
\end{figure}

\customparagraph{Q1. To what extent can upsampling be reduced?}
Initially, we investigated the required number of upsampling steps $L$.
The results are presented in Table~\ref{tab:analysis_upsampling}.
We found that \textit{FVG$+$VPFD$_0$} significantly degrades DNSMOS and SECS because $\mathcal{V}_0^{\textnormal{feat}}$ consists of only a single convolution layer, which is inadequate for extracting effective features.
In contrast, \textit{FVG$+$VPFD$_L$} with $L \geq 1$ achieved UTMOS, DNSMOS, CER, and SECS comparable to those of \textit{FVG}.
Notably, \textit{FVG$+$VPFD$_1$} achieved this while reducing the training time and memory consumption by factors of 9.6 and 11.4, respectively.
To investigate why one upsampling is sufficient, we visualized the vocoder features in Figure~\ref{fig:intfeat}.
As shown in (c), periodic structures, which are crucial for waveform representation but absent in the mel-spectrogram (a) and features without upsampling (b), emerge with a single upsampling, suggesting its effectiveness.
Based on these findings, we set $L = 1$ for subsequent experiments.

\begin{table}[h]
  \vspace{-2mm}
  \caption{Comparison of performance with varying numbers of upsampling steps.
    The scores improve as the color transitions from blue to white to red.}
  \vspace{-2mm}
  \label{tab:analysis_upsampling}
  \newcommand{\spm}[1]{{\tiny$\pm$#1}}
  \setlength{\tabcolsep}{1pt}
  \centering
  \scriptsize{
    \begin{tabularx}{\columnwidth}{cCCCCCC}
      \toprule
      \texttt{Model}
      & \texttt{UTMOS$\uparrow$} & \texttt{DNSMOS$\uparrow$} & \texttt{CER$\downarrow$} & \texttt{SECS$\uparrow$} & \texttt{Time$\downarrow$} & \texttt{Memory$\downarrow$}
      \\ \midrule
      FVG
      & 3.96 & 3.77 & 1.3 & 0.847 & \worstp{47.0} & \worstpp{66.3}
      \\ \midrule
      FVG$+$VPFD$_0$
      & \goodm{3.98} & \worst{3.66} & 1.3 & \worst{0.843} & \better{~~1.5} &  \better{~~2.8}
      \\
      FVG$+$VPFD$_1$
      & \good{3.99} & \good{3.79} & \goodm{1.2} & \good{0.851} & \good{~~4.9} &  \good{~~5.8}
      \\
      FVG$+$VPFD$_2$
      & \good{3.99} & \good{3.79} & \goodm{1.2} & \goodm{0.850} & \bad{14.1} & \bad{17.9}
      \\
      FVG$+$VPFD$_3$
      & \goodm{3.98} & \goodm{3.78} & \goodm{1.2} & 0.849 & \worse{23.0} & \worse{31.1}
      \\
      FVG$+$VPFD$_4$
      & \goodm{3.98} & \goodm{3.78} & 1.3 & 0.849 & \worst{31.8} & \worst{44.3}
      \\ \bottomrule
    \end{tabularx}
  }
  \vspace{-6mm}
\end{table}

\smallskip
\customparagraph{Q2. How should the vocoder feature extractor $\mathcal{V}_1^{\textnormal{feat}}$ be handled during training?}
Next, we investigated the handling of $\mathcal{V}_1^{\textnormal{feat}}$ during the training.
In particular, we examined the effects of pretraining and freezing parameters.
The results in Table~\ref{tab:analysis_feature_extractor} demonstrate that both strategies are crucial for all metrics.
Based on these findings, we employed pretrained and frozen $\mathcal{V}_1^{\textnormal{feat}}$ in subsequent experiments.

\begin{table}[h]
  \vspace{-2mm}
  \caption{Analysis of the importance of pretraining and freezing the vocoder feature extractor.}
  \vspace{-2mm}
  \label{tab:analysis_feature_extractor}
  \newcommand{\spm}[1]{{\tiny$\pm$#1}}
  \setlength{\tabcolsep}{1pt}
  \centering
  \scriptsize{
    \begin{tabularx}{\columnwidth}{ccCCCCCC}
      \toprule
      \texttt{Pretrained} & \texttt{Frozen}
      & \texttt{UTMOS$\uparrow$} & \texttt{DNSMOS$\uparrow$} & \texttt{CER$\downarrow$} & \texttt{SECS$\uparrow$} & \texttt{Time$\downarrow$} & \texttt{Memory$\downarrow$}
      \\ \midrule
                          &
      & \worst{3.62} & \worst{3.54} & 1.3  & \worst{0.814} & ~~4.9 &  ~~5.8
      \\
      \checkmark &
      & \worse{3.74} & \worse{3.64} & 1.3 & \worse{0.837} & ~~4.9 &  ~~5.8
      \\
                          & \checkmark
      & \bad{3.97} & \bad{3.68} & 1.3 & \bad{0.843} & ~~4.9 &  ~~5.8
      \\
      \checkmark & \checkmark
      & \good{3.99} & \good{3.79} & \goodm{1.2} & \good{0.851} & ~~4.9 &  ~~5.8
      \\ \bottomrule
    \end{tabularx}
  }
  \vspace{-4mm}
\end{table}

\subsection{Comparative study}
\label{subsec:comparative_study}

\customparagraph{Comparison with other training acceleration techniques.}
This study accelerated training by reducing the upsampling steps in the discriminator.
To evaluate this approach, we compared it with several alternatives:
(1) \textit{FVG$_{\text{early}}$}, which reduces the number of training epochs from 100 to 10, thereby matching the training time of FVG$+$VPFD$_1$.
(2) \textit{FVG w/o MRD}, which ablates MRD in VWD.
(3) \textit{FVG w/o MPD}, which ablates MPD in VWD.
(4) \textit{FVG$+$MelD$_{\text{small}}$}, which replaces VWD with a mel-spectrogram discriminator (MelD) with an architecture similar to that of MRD. 
(5) \textit{FVG$+$MelD$_{\text{large}}$}, which increases the number of channels in \textit{FVG$+$MelD$_{\text{small}}$} to match the training time of FVG$+$VPFD$_1$.
The results in Table~\ref{tab:comparison_with_alternatives} indicate the following:
(1) The performance decreases as the number of training epochs decreases.
(2) and (3) Ablating the discriminator in VWD degrades the performance and does not significantly reduce the training time.
(4) and (5) \textit{FVG$+$MelD} fails to achieve a high DNSMOS, regardless of the model size.
These results indicate that \textit{FVG$+$VPFD$_1$} is most effective for reducing training time while maintaining VC performance.

\begin{table}[h]
  \vspace{1mm}
  \caption{Comparison with other training acceleration techniques.
    The scores improve as the color transitions from blue to white to red.}
  \vspace{-2mm}
  \label{tab:comparison_with_alternatives}
  \newcommand{\spm}[1]{{\tiny$\pm$#1}}
  \setlength{\tabcolsep}{0.5pt}
  \centering
  \scriptsize{
    \begin{tabularx}{\columnwidth}{ccCCCCCC}
      \toprule
      & \texttt{Model}
      & \texttt{UTMOS$\uparrow$} & \texttt{DNSMOS$\uparrow$} & \texttt{CER$\downarrow$} & \texttt{SECS$\uparrow$} & \texttt{Time$\downarrow$} & \texttt{Memory$\downarrow$}
      \\ \midrule
      & FVG
      & 3.96 & 3.77 & 1.3 & 0.847 & \worstp{47.0} & \worstpp{66.3}
      \\ \midrule
      (1) & FVG$_{\text{early}}$
      & \worst{3.82} & \worst{3.72} & 1.3 & \worst{0.843} & \goodp{~~4.7} & \worstpp{66.3}
      \\
      (2) & FVG w/o MRD
      & \badm{3.95} & \badm{3.75} & \goodm{1.2} & \bad{0.845} & \worst{32.1} & \worst{43.6}
      \\
      (3) & FVG w/o MPD
      & \worst{3.87} & \worst{3.70} & 1.3 & 0.847 & \worstp{40.3} & \worstp{56.7}
      \\
      (4) & FVG$+$MelD$_\text{small}$
      & \goodm{3.98} & \worst{3.66} & 1.3 & \bad{0.845} & \better{~~1.9} & \better{~~3.1}
      \\
      (5) & FVG$+$MelD$_\text{large}$
      & \good{3.99} & \worst{3.68} & 1.3 & \bad{0.845} & \good{~~4.9} & \goodm{~~6.1}
      \\ \midrule
      & FVG$+$VPFD$_1$
      & \good{3.99} & \good{3.79} & \goodm{1.2} & \good{0.851} & \good{~~4.9} & \good{~~5.8}
      \\ \bottomrule
    \end{tabularx}
  }
  \vspace{-2mm}
\end{table}

\smallskip
\customparagraph{Subjective evaluation.}
For a comprehensive evaluation, we conducted MOS tests for 90 different speaker/sentence pairs to evaluate speech quality (\textit{qMOS} on a five-point scale: 1 = bad, 2 = poor, 3 = fair, 4 = good, and 5 = excellent) and speaker similarity (\textit{sMOS} on a four-point scale: 1 = different (sure), 2 = different (not sure), 3 = same (not sure), and 4 = same (sure)).
In both tests, we compared \textit{FVG$+$VPFD$_1$} with \textit{FVG} to assess the effect of replacing VWD with \textit{VPFD}.
Ground-truth speech and speech converted using DiffVC-30~\cite{VPopovICLR2022}, a widely used baseline, were included as anchor samples.
For each test, more than 1,000 responses were collected from 11 participants.
As shown in Table~\ref{tab:subjective_evaluation}, \textit{FVG$+$VPFD$_1$} achieved performance comparable to \textit{FVG} in both objective and subjective evaluations.

\begin{table}[h]
  \vspace{-1mm}
  \caption{Subjective evaluations with 95\% confidence intervals.
    Objective scores are also included as a reference.}
  \vspace{-2mm}
  \label{tab:subjective_evaluation}
  \newcommand{\spm}[1]{{\tiny$\pm$#1}}
  \setlength{\tabcolsep}{1pt}
  \centering
  \scriptsize{
    \begin{tabularx}{\columnwidth}{cCCCCCC}
      \toprule
      \texttt{Model}
      & \texttt{qMOS$\uparrow$} & \texttt{sMOS$\uparrow$} & \texttt{UTMOS$\uparrow$} & \texttt{DNSMOS$\uparrow$} & \texttt{CER$\downarrow$} & \texttt{SECS$\uparrow$}
      \\ \midrule
      Ground truth
      & 4.43\spm{0.08} & 3.57\spm{0.08} & 4.14 & 3.75 & 0.1 & 0.871
      \\ \midrule
      DiffVC-30
      & 3.60\spm{0.10} & \bad{2.37\spm{0.12}} & 3.76 & \badm{3.75} & \worst{5.4} & \worst{0.802}
      \\ \midrule
      FVG
      & \goodm{3.61\spm{0.09}} & \goodm{2.64\spm{0.12}} & 3.96 & 3.77 & 1.3 & 0.847
      \\
      FVG$+$VPFD$_1$
      & \good{3.63\spm{0.10}} & \good{2.69\spm{0.12}} & \good{3.99} & \good{3.79} & \goodm{1.2} & \good{0.851}
      \\ \bottomrule
    \end{tabularx}
  }
  \vspace{-2mm}
\end{table}

\subsection{Generalizability analysis}
\label{subsec:generalizability_analysis}

To investigate dataset dependency, we also conducted experiments using the LibriTTS dataset~\cite{HZenIS2019}.
The results presented in Table~\ref{tab:analysis_libritts} demonstrate a similar tendency as in the VCTK dataset; that is, \textit{FVG$+$VPFD$_1$} achieves VC performance comparable to that of \textit{FVG} while reducing both the training time and memory consumption by factors of 9.6 and 11.4, respectively.

\begin{table}[h]
  \vspace{-1mm}
  \caption{Results on LibriTTS dataset.
    $^\dag$Ground-truth converted speech does not necessarily exist in LibriTTS.
    Therefore, source speech was used to calculate the scores.}
  \vspace{-2mm}
  \label{tab:analysis_libritts}
  \newcommand{\spm}[1]{{\tiny$\pm$#1}}
  \setlength{\tabcolsep}{1pt}
  \centering
  \scriptsize{
    \begin{tabularx}{\columnwidth}{cCCCCCC}
      \toprule
      \texttt{Model}
      & \texttt{UTMOS$\uparrow$} & \texttt{DNSMOS$\uparrow$} & \texttt{CER$\downarrow$} & \texttt{SECS$\uparrow$} & \texttt{Time$\downarrow$} & \texttt{Memory$\downarrow$}
      \\ \midrule
      Ground truth$^\dag$
      & 4.06 & 3.70 & 0.6 & -- & -- & --
      \\ \midrule
      FVG
      & 3.94 & 3.75 & 1.2 & 0.843 & \worst{176.3} & \worstpp{66.3}
      \\
      FVG$+$VPFD$_1$
      & \good{4.06} & \goodm{3.76} & \goodm{1.1} & \good{0.847} & \good{~~18.4} & \good{~~5.8}
      \\ \bottomrule
    \end{tabularx}
  }
  \vspace{-2mm}
\end{table}

\section{Conclusion}
\label{sec:conclusion}

We proposed \textit{VPFD}, which uses vocoder features for adversarial training.
The results show that a pretrained and frozen vocoder feature extractor with a single upsampling step is necessary and sufficient to achieve comparable VC performance while significantly reducing training time and memory usage.
Given the broad use of adversarial training with acoustic features in TTS and VC, we see potential for applying this approach to other tasks in future research.

\bibliographystyle{IEEEtran}
\bibliography{refs}

\end{document}